\documentclass[lettersize,journal]{IEEEtran}
\usepackage{amsmath,amsfonts}
\usepackage{algorithmic}
\usepackage{algorithm}
\usepackage{array}
\usepackage[caption=false,font=normalsize,labelfont=sf,textfont=sf]{subfig}
\usepackage{textcomp}
\usepackage{stfloats}
\usepackage{url}
\usepackage{verbatim}
\usepackage{graphicx}
\usepackage{epstopdf}
\usepackage{cite}
\usepackage{bm}
\usepackage{multirow}
\usepackage{setspace}
\usepackage{booktabs}
\begin{document}

\title{NLOS Error Mitigation Using Weighted Least Squares and Kalman Filter in UWB Positioning}

\author{Ruixin Fan and Xin Du
\thanks{Ruixin Fan and Xin Du are with the  College of Information Science \& Electronic Engineering, Zhejiang University, China. (E-mail: fanruixin@zju.edu.cn; duxin@zju.edu.cn)}%
}



\maketitle

\begin{abstract}
In wireless positioning systems, non-line-of-sight (NLOS) is a challenging problem. 
NLOS causes great ranging bias and location error, so NLOS mitigation is essential for high accuracy positioning. 
In this letter, we propose the Weighted-Least-Squares Robust Kalman Filter (WLS-RKF) for NLOS identification and mitigation.
WLS-RKF employs a hypothesis test based on Mahalanobis distance for NLOS identification, and updates the corresponding Kalman filter using the WLS solution.
It requires no prior knowledge about NLOS distribution or signal features. 
We perform simulations and experiments for ultra-wideband (UWB) positioning in various scenarios. 
The results confirm that WLS-RKF effectively mitigates NLOS error and achieves 5cm positioning accuracy. 
\end{abstract}


\section{Introduction}
\IEEEPARstart{W}{ireless} positioning has been widely used for robot navigation and other scenarios. For indoor positioning where the Global Navigation Satellite System (GNSS) is unavilable, ultra-wideband (UWB) is often adopted for its high accuracy and low cost. 
In a typical UWB positioning system, the distances between the tag and the anchors can be measured based on time of flight (TOF), and then the tag's position can be estimated. 
However, if the signal is obstructed, it propagates slower in the obstacle than in the air or arrives at the receiver from reflections, which makes the measured distance larger than the true distance \cite{Jourdan2008}. 
This is referred to as the non-line-of-sight (NLOS) effect, which leads to a significant location error. 

Currently, there are two kinds of approaches for NLOS mitigation. 
The first kind is to identify NLOS measurements based on the features of the received signal, such as confidence metric, delay spread, and change of SNR \cite{schroeder2007nlos}. 
\cite{Musa2019} proposes a decision tree-based method using the UWB channel quality indicators. 
After NLOS identification, an accurate position is calculated by the weighted-least-squares (WLS) technique \cite{guvencc2007nlos}. 
Because the signal features highly depend on the propagation environment, such algorithms are difficult to be widely used \cite{wang2018robust}.
The second kind is based on the measured distances.  
For example, the residual-weighting (Rwgh) algorithm adopts distance residual as a measure of the presence of NLOS\cite{chen1999non}.
Algorithms such as RCC-Rwgh \cite{jiao2009performance} are proposed to reduce the computational complexity of Rwgh.
However, \cite{liu2013analysis} shows that these methods work only for location estimates within the convex-hull region enclosed by the anchors.
Based on the assumption that the measured distance is positively biased when NLOS occurs,
\cite{le2003mobile} uses the standard deviation of measured distances for NLOS identification.
Recently, \cite{li2019comparative} proposes the robust Kalman Filter (RKF). RKF uses the Mahalanobis distance to detect outliers of the measurements, which is probably affected by NLOS.
Then the effect on the posterior estimation can be mitigated by increasing the covariance of the filter.  
However, we find that the positioning error increases when the tag moves in the NLOS condition.

In this letter, we focus mainly on the second kind of approaches.
We propose the Weighted-Least-Squares Robust Kalman Filter (WLS-RKF) method for NLOS mitigation.
In WLS-RKF, the NLOS is identified by a hypothesis test based on Mahalanobis distance. Then WLS technique is adopted where the NLOS measurement gets a smaller weight, and the corresponding filter is updated using the WLS solution.
It uses only the measured distances and requires no prior knowledge about NLOS distribution or signal features. 
We evaluate the method by simulations and experiments in UWB positioning. 
The results show that the location error is reduced by over 90\% and positioning accuracy is up to 5cm.

The rest of this letter is organized as follows. In Section II, we give a brief introduction of wireless positioning model, the Kalman filter and the NLOS error model. In Section III, the WLS-RKF is presented and discussed in detail. The simulation and experiment results for the scenario of UWB positioning are shown in Section IV and V respectively. Finally, Section IV gives a conclusion.

\section{Preliminaries}

\subsection{Wireless Positioning Model}
We consider a wireless positioning system with $N$ anchors $A_1$, $A_2$, ..., $A_N$ and a tag $T$. 
The position vector of $A_i$ is $\bm{Z}_i = [x_i\ y_i]^T$, and the position vector of $T$ is $\bm{Z} = [x\ y]^T$. 
$r_i$ is the measured distance between $T$ and $A_i$. 
Once all the distances are available, we have the following equations:
\begin{equation}
	\label{localization equation}
	||\bm{Z} - \bm{Z}_i|| = \sqrt{(x-x_i)^2+(y-y_i)^2} = r_i, \ i=1,2,...,N.
\end{equation}

Let $w_i$ denote the weight of the $i$th equation in (\ref{localization equation}).
Then the WLS estimate of the tag's position can be expressed as
\begin{equation}
	\label{WLS estimate}
	\hat{\bm{Z}} = \arg \min_{\bm{Z}} \sum_{i=1}^{N} w_i^2 (r_i - ||\bm{Z} - \bm{Z}_i||)^2,
\end{equation}
which can be solved by Gauss-Newton techniques \cite{gustafsson2005mobile} or other numerical search methods.

\subsection{Kalman Filter}
The Kalman filter is used to smooth the measurement data. From \cite{le2003mobile}, the state data vector satisfies:
\begin{equation}
	\bm{X}_{k} = \bm{A} \bm{X}_{k-1} + \bm{\Gamma} \bm{W}_{k-1},
\end{equation}
where $\bm{X}_k$ is the state vector at the time $t_k$. $\bm{W}_{k-1}$ is the driving noise vector with covariance matrix $\bm{Q} = \sigma_u^2 \bm{I}$ and
\begin{equation}
	\bm{A} = \left[
	\begin{array}{cc}
		\bm{I} & \Delta t \bm{I} \\
		\bm{O} & \bm{I}
	\end{array}
	\right], \quad
	\bm{\Gamma}= \left[
	\begin{array}{c}
		\bm{O} \\
		\Delta t \bm{I}
	\end{array}
	\right].
\end{equation}

The measurement process is
\begin{equation}
	\bm{Y}_k = \bm{H} \bm{X}_{k} + \bm{U}_k,
\end{equation}
where $\bm{Y}_k$ is the measured data vector, $\bm{H} = [\bm{I} \ \bm{O}]$, and $\bm{U}_k$ is the measurement noise vector with covariance matrix $\bm{R} = \sigma_x^2 \bm{I}$.

In each iteration, the Kalman filter includes the following steps:
\begin{align}
	\hat{\bm{X}}_{k,k-1} &= \bm{A} \hat{\bm{X}}_{k-1,k-1} \label{prediction}\\
	\bm{P}_{k,k-1} &= \bm{A} \bm{P}_{k-1,k-1} \bm{A}^T + \bm{\Gamma} \bm{Q} \bm{\Gamma}^T \label{kalman2}\\
	\bm{K}_k &= \bm{P}_{k,k-1} \bm{H}^T (\bm{H} \bm{P}_{k,k-1} \bm{H}^T + \bm{R}) \\
	\hat{\bm{X}}_{k,k} &= \hat{\bm{X}}_{k,k-1} + \bm{K}_k(\bm{Y}_k - \bm{H} \hat{\bm{X}}_{k,k-1}) \label{kalman4}\\
	\bm{P}_{k,k} &= (\bm{I} - \bm{K}_k \bm{H}) \bm{P}_{k,k-1} \label{kalman5}
\end{align}
where $\bm{K}_k$ is the Kalman gain and $\bm{P}_{k,k}$ is the covariance matrix of $\hat{\bm{X}}_{k,k}$.

In wireless positioning, Kalman filter can be used to smooth distance measurement data. The state vector is $\bm{X}_k = [r_k\ \dot{r}_k]^T$, where $\dot{r}_k$ denotes the first derivative of $r_k$.

\subsection{NLOS Ranging Model}
Due to the NLOS effect, distance measurement is positively biased. 
The measured distance $r$ can be expressed as \cite{Jourdan2008}:
\begin{equation}
	\label{disitance model}
	r = d + b + \epsilon,
\end{equation}
where $d$ is the true distance, $b$ is the ranging bias caused by NLOS propagation, and $\epsilon$ is the measurement noise.

Measurement noise can be modeled as Gaussian distributed $\epsilon \sim N(0, \sigma_m)$. 
NLOS bias can be modeled in different ways such as exponentially distributed, uniformly distributed and Gaussian distributed \cite{guvencc2007nlos}. 
In \cite{silva2020ranging}, a simplified through-the-wall (TTW) NLOS ranging model is proposed. When an RF wave traverses a wall, the ranging bias is
\begin{equation}
	\label{TTW model}
	b \approx w(\sqrt{\epsilon_r}-1) + 0.31w \theta_i^2,
\end{equation}
where $\epsilon_r$ is the wall's real relative permittivity, $w$ is the wall’s thickness, and $\theta_i$ is the incidence angle. 

In general, the NLOS bias is much larger than the measurement noise, that is, $b >> \epsilon$. Data from UWB range measurements shows that $\sigma_m$ is only a few centimeters\cite{gonzalez2009mobile}, while $b$ can be up to 1.5m \cite{Jourdan2005}.

\section{Weighted-Least-Squares \\ Robust Kalman Filter}
In this section we detail the proposed algorithm, Weighted-Least-Squares Robust Kalman Filter (WLS-RKF).  The main idea is to identify the positive bias of the NLOS measurements, and mitigate the position error using the other line-of-sight (LOS) measurements.

\subsection{NLOS Identification}
\label{NLOS identification}
Kalman filter can be used to identify outliers in the stream of data. 
Similar to \cite{li2019comparative},
we use a Kalman filter KF$_i$ to handle the distance measurements to the $i$th anchor respectively.
In each iteration, the Kalman filter predicts the distance ${d}_{ki} = \bm{H} \hat{\bm{X}}_{k,k-1}$. In LOS condition, the measured distance $\bm{Y}_k = r_{ki}$ obeys the Gaussian distribution, with the mean of ${d}_{ki}$ and the variance of ($\bm{H} \bm{P}_{k,k-1} \bm{H} + \bm{R}$). 
Therefore, the square of the Mahalanobis distance from $r_{ki}$ to ${d}_{ki}$ obeys the $\chi^2$ distribution\cite{chang2014kalman} with the degree of freedom of 1, that is 
\begin{equation}
	\label{Mahalanobis}
	\gamma_{ki} = (r_{ki} - {d}_{ki})^T (\bm{H} \bm{P}_{k,k-1} \bm{H} + \bm{R})^{-1} (r_{ki} - {d}_{ki}) \sim \chi_1^2,
\end{equation}

For the significance level $\alpha$, we have
\begin{equation}
	\text{Pr}(\gamma_{ki} \leq \chi_{1,\alpha}^2) = 1 - \alpha,
\end{equation}
where Pr() is the probability of a random event, and $\chi_{1,\alpha}^2$ is the $\alpha$-quantile of the $\chi_1^2$ distribution. In this letter, we choose $\alpha = 0.001$, and accordingly, $\chi_{1,\alpha}^2 = 6.2$. 

If the measured distance does not satisfy the condition $\gamma_{ki} \leq \chi_{1,\alpha}^2$, it is considered as an outlier, which is probably caused by NLOS. 
Since the NLOS bias is always positive, we add another constraint $r_{ki} > {d}_{ki}$, which can reduce the probability of false alarm.
Therefore, the following hypothesis test is employed to identify NLOS measurements:
\begin{equation}
	\label{identification}
	\begin{aligned}
		&H_0: \gamma_{ki} > \chi_{1,\alpha}^2\ \text{and}\ r_{ki} > {d}_{ki} \quad \text{NLOS case} \\
		&H_1: \gamma_{ki} \leq \chi_{1,\alpha}^2\ \text{ or }\ r_{ki} \leq {d}_{ki}  \quad \text{LOS case}
	\end{aligned}
\end{equation}

\subsection{NLOS Mitigation}
After NLOS measurements are identified, we use the WLS technique to mitigate the position error. According to (\ref{WLS estimate}), the weights and the distances from the tag to the anchors should be determined.

The distance $\hat{r}_{ki}$ is the output of the $i$th Kalman filter:
\begin{equation}
	\label{weighting function}
	\hat{r}_{ki} = \left\{  
	\begin{array}{cc}
		{d}_{ki},\quad &\text{NLOS case} \\
		\bm{H} \hat{\bm{X}}_{k,k},\quad &\text{ LOS case } \\
	\end{array}
	\right.
\end{equation}
where ${d}_{ki}$ denotes the predicted distance, and $\hat{\bm{X}}_{k,k}$ is the estimated state vector by (\ref{kalman4}). $\bm{H} \hat{\bm{X}}_{k,k}$ denotes the first element of $\hat{\bm{X}}_{k,k}$, which is the estimated distance.
In LOS case, the Kalman filter gives an accurate estimation of the distance.
In NLOS case, however, the measured distance ${r}_{ki}$ is biased and not reliable. So we use the predicted distance.

The weight $w_i$ is selected as:
\begin{equation}
	\label{weighting function}
	w_i = \left\{  
	\begin{array}{cc}
		\dfrac{1}{\sqrt{{\gamma_{ki}}/{\chi_{1,\alpha}^2}} },\quad &\text{NLOS case} \\
		1,\quad &\text{ LOS case } \\
	\end{array}
	\right.
\end{equation}
where $\sqrt{{\gamma_{ki}}/{\chi_{1,\alpha}^2}}$ is the Mahalanobis distance normalized with the threshold $\chi_{1,\alpha}^2$.

In NLOS case, according to (\ref{identification}),  $w_i$ is less than 1 and inversely proportional to the Mahalanobis distance, which is determined by the NLOS bias. 
Therefore, distance measurement with a large NLOS bias will get a small weight. 
The WLS solution is mainly determined by LOS measurements, and the location error caused by NLOS bias is mitigated.

Then the estimated position of the tag $\hat{\bm{Z}}_k$ is the WLS solution:
\begin{equation}
	\hat{\bm{Z}}_k = \mathop{\arg\min}\limits_{\bm{Z}} \sum_{i=1}^{N} w_i^2 (\hat{r}_{ki} - ||\bm{Z} - \bm{Z}_i||)^2.
\end{equation}

In LOS case, we update the state of Kalman filter with the measured distance $\bm{Y}_k = r_{ki}$.
In NLOS case, however, this will make the filter will enter the wrong state, since the measured distance $r_{ki}$ is biased. 
Therefore, after we get $\hat{\bm{Z}}_k$, we update the state of Kalman filter with the distance between $\hat{\bm{Z}}_k$ and the anchor's position $\bm{Z}_i$:
\begin{equation}
	\bm{Y}_k = ||\hat{\bm{Z}}_k - \bm{Z}_i||.
\end{equation}

As discussed before, $\hat{\bm{Z}}_k$ is mainly determined by the LOS measurements. So the state of the Kalman filter can be better estimated with the information of the LOS measurements. This ensures that the filter gives an accurate distance prediction for NLOS identification in the next iteration.
By comparison, in RKF \cite{li2019comparative}, distance measurements to each anchor are handled with an independent Kalman filter. Although the efect of NLOS can be mitigated by increasing the covariance, the state of the filter still diverges slowly.

\begin{algorithm}[t]
	\caption{WLS-RKF}
	
	{\bf Input:}
	number of anchors $N$, the threshold $\chi_{1,\alpha}^2$, measured distances $r_{ki}$ at time $t_k$, $i = 1, 2, ..., N, k = 1, 2, ...$\\
	{\bf Output:}
	estimated positions $\hat{\bm{Z}}_k$ at time $t_k$, $k = 1, 2, ...$
	
	\begin{algorithmic}[1]
		\STATE Initialization for Kalman filters KF$_i$, $i = 1, 2, ..., N$
		\FOR{k = 1, 2, ...}
			\FOR{$i = 1, 2, ..., N$} 
				\STATE Predict distance ${d}_{ki}$ by KF$_i$ \label{predict}
				\STATE Calculate $\gamma_{ki}$ by (\ref{Mahalanobis})
				\setstretch{1.25}
				\IF{$\gamma_{ki} > \chi_{1,\alpha}^2$ \AND $r_{ki} > {d}_{ki}$} \label{test1}
					\STATE $NLOS_i$ = \TRUE
					\STATE $\hat{r}_{ki} = {d}_{ki}$  \label{NLOS1}
					\STATE $w_i = \sqrt{{\chi_{1,\alpha}^2} / {\gamma_{ki}}}$ \label{NLOS2}
					\setstretch{1.5}
				\ELSE
					\setstretch{1}
					\STATE $NLOS_i$ = \FALSE
					\STATE Update state for KF$_i$ with $\bm{Y}_k = r_{ki}$ \label{LOS1}
					\STATE $\hat{r}_{ki} = \bm{H} \hat{\bm{X}}_{k,k}$ \label{LOS2}
					\STATE $w_i = 1$ \label{LOS3}
				\ENDIF
			\ENDFOR
			
			\STATE $\hat{\bm{Z}}_k = \mathop{\arg\min}\limits_{\bm{Z}} \sum_{i=1}^{N} w_i^2 (\hat{r}_{ki} - ||\bm{Z} - \bm{Z}_i||)^2$
			\setstretch{1.5}
			
			\FOR{$i = 1, 2, ..., N$} 
				\setstretch{1}
				\IF{$NLOS_i$ == \TRUE}
					\STATE Update state for KF$_i$ with $\bm{Y}_k = ||\hat{\bm{Z}}_k - \bm{Z}_i||$ \label{NLOS3}
				\ENDIF
			\ENDFOR
			\setstretch{1}
		\ENDFOR
	\end{algorithmic}
\end{algorithm}

The operation of the proposed WLS-RKF is summarized in Algorithm 1. For each distance measurement $r_{ki}$, LOS/NLOS case is identified in Line \ref{test1}. The NLOS case is handled in Line \ref{NLOS1}-\ref{NLOS2} and \ref{NLOS3}. The LOS case is handled in Line \ref{LOS1}-\ref{LOS3}. 

It should be noted that enough LOS measurements is required to ensure the position accuracy. For 2-D positioning, the number of LOS measurements should be no less than 2. We suggest that the positioning system includes at least 4 anchors to achieve a better performance.

\section{Simulations and Results}
\subsection{Simulation Setup}
\begin{figure}[t]
	\centering
	\subfloat[]{\includegraphics[width=4cm]{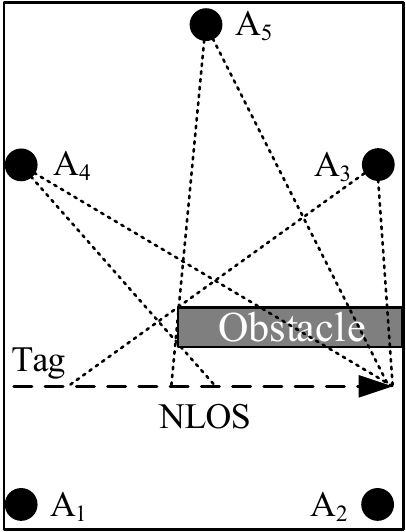}}
	\hfil
	\subfloat[]{\includegraphics[width=4cm]{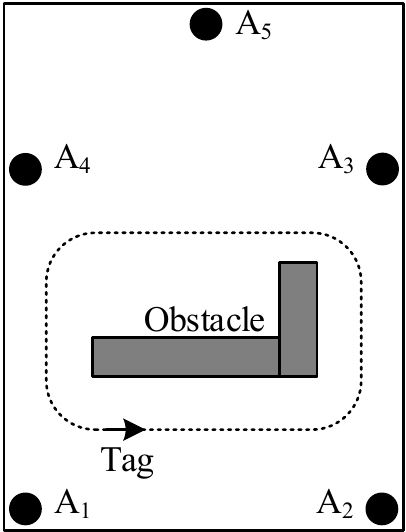}}
	\caption{Simulation cases. (a) Case 1 \& 2. (b) Case 3 \& 4. The anchor $A_5$ is excluded in Case 1 and 3. The obstacle will introduce NLOS bias to the distance measurements.}
	\label{simulation}
\end{figure}

We consider 4 representative cases, as shown in Fig.\ref{simulation}. 
\begin{itemize}
	\item Case 1: There are 4 anchors with coordinates $A_1$(0, 0), $A_2$(10, 0), $A_3$(10, 10) and $A_4$(0, 10). The tag $T$ moves from $(0, 3)$ to $(10, 3)$ in the straight line at a constant velocity of 0.5m/s.
	\item Case 2: The settings are the same as case 1, and another anchor with coordinate $A_5$(5, 15) is added.
	\item Case 3: There are 4 anchors with the same coordinates as in case 1. The tag $T$ moves around the rectangle at a constant velocity of 0.5m/s for two laps. The rectangle is 8m long and 6m wide, and its corner is round with a radius of 0.5m.
	\item Case 4: The settings are the same as case 3, and another anchor with coordinate $A_5$(5, 15) is added.
\end{itemize}

In case 1 and 2, there is a wall with the length $L$ and width $W$. The wall makes the path $TA_3$, $TA_5$ and $TA_4$ change from LOS to NLOS successively as $T$ moves along the line.
In case 2 and 4, two walls with the length $L_1$, $L_2$ and width $W$ is placed orthogonally. The walls introduce a more severe obstruction: at least one path $TA_i$ is NLOS at almost any point of the trajectory. $L$, $L_1$, $L_2$ and $W$ are randomly chosen from 3-8m, 4-7m, 2-5m and 0.3-0.7m. The settings ensure that at least 2 LOS measurements exist at any time.

To create the measurement data for simulation, we calculate the true distances from $T$ to $A_i$ with sampling interval 0.05s and add the measurement noise and NLOS bias to the true distances. 
The measurement noise is Gaussian distribution with standard deviation $\sigma_m = 0.02m$ and the NLOS bias is calculated with the TTW model \ref{TTW model}, where $\epsilon_r = 6$ and $\theta_i$ is calculated from the geometry of $TA_i$ and the walls.

For the proposed WLS-RKF, we select $\chi_{1,\alpha}^2$=6.2. 
The parameters of the Kalman filters are $\sigma_x = \sigma_m = 0.02$ and $\sigma_u = 0.5$. 
And the initial value of $\bm{P}$ is $\bm{P}_0 = diag(\sigma_m^2, 0)$. 
We also simulate with the RKF \cite{li2019comparative} for comparison, and with unweighted LS as the benchmark. 

\subsection{Simulation Results}
\begin{figure}[!tb]
	\centering
	\subfloat[]{\includegraphics[width=8.5cm]{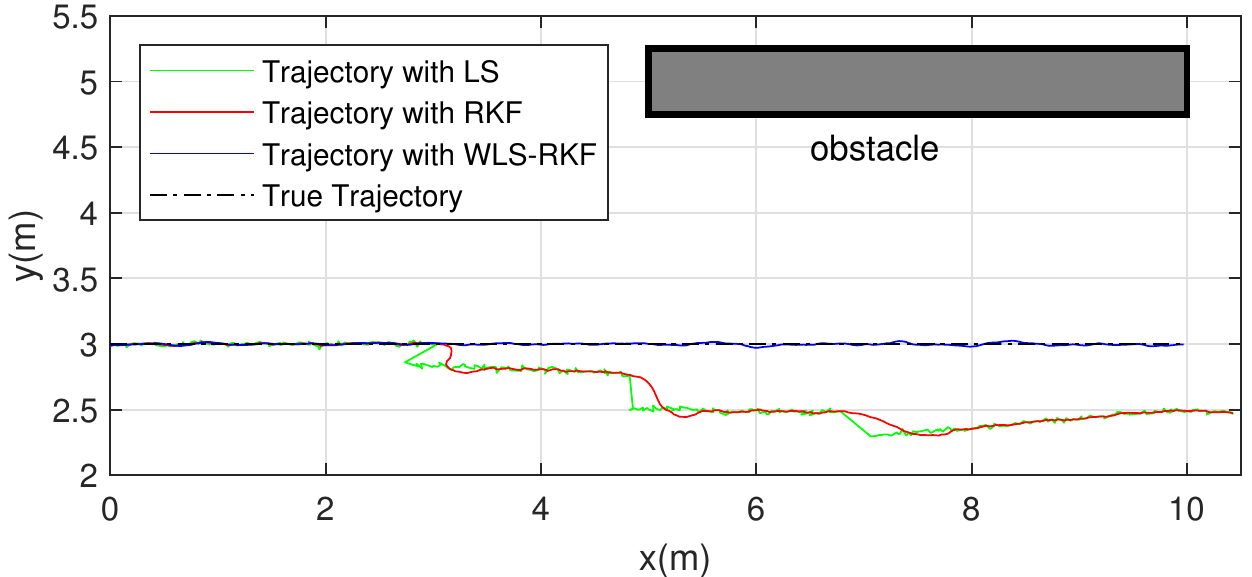}}
	\vspace{0mm}
	\subfloat[]{\includegraphics[width=8.5cm]{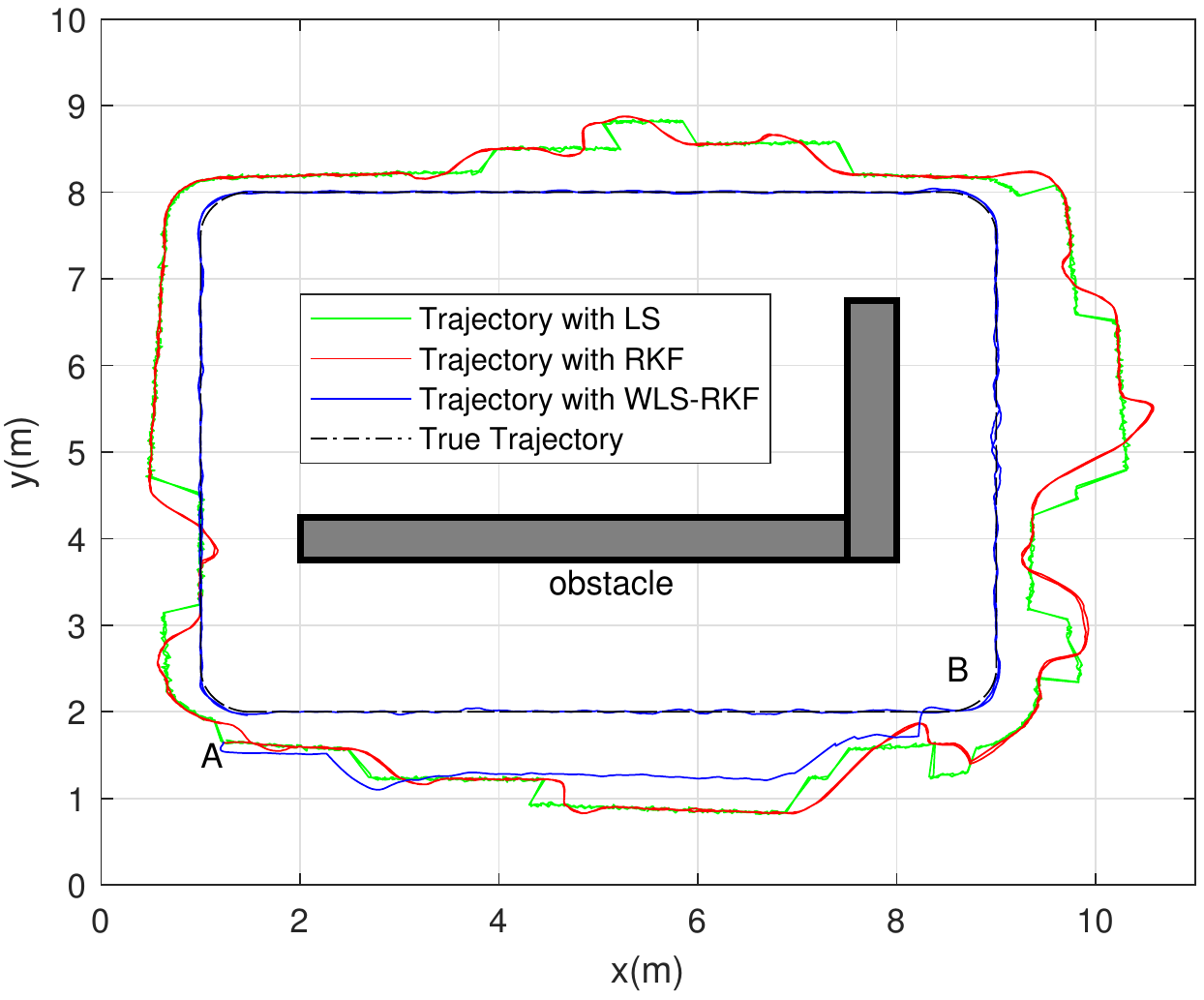}}
	\caption{Simulation results for LS, RKF and the proposed WLS-RKF. (a) Case 2. (b) Case 4. The results of Case 1 and 3 are similar to (a) and (b).}
	\label{simulation cases}
\end{figure}

\begin{figure}[tb]
	\centering
	\subfloat[]{\includegraphics[width=8cm, height=5cm]{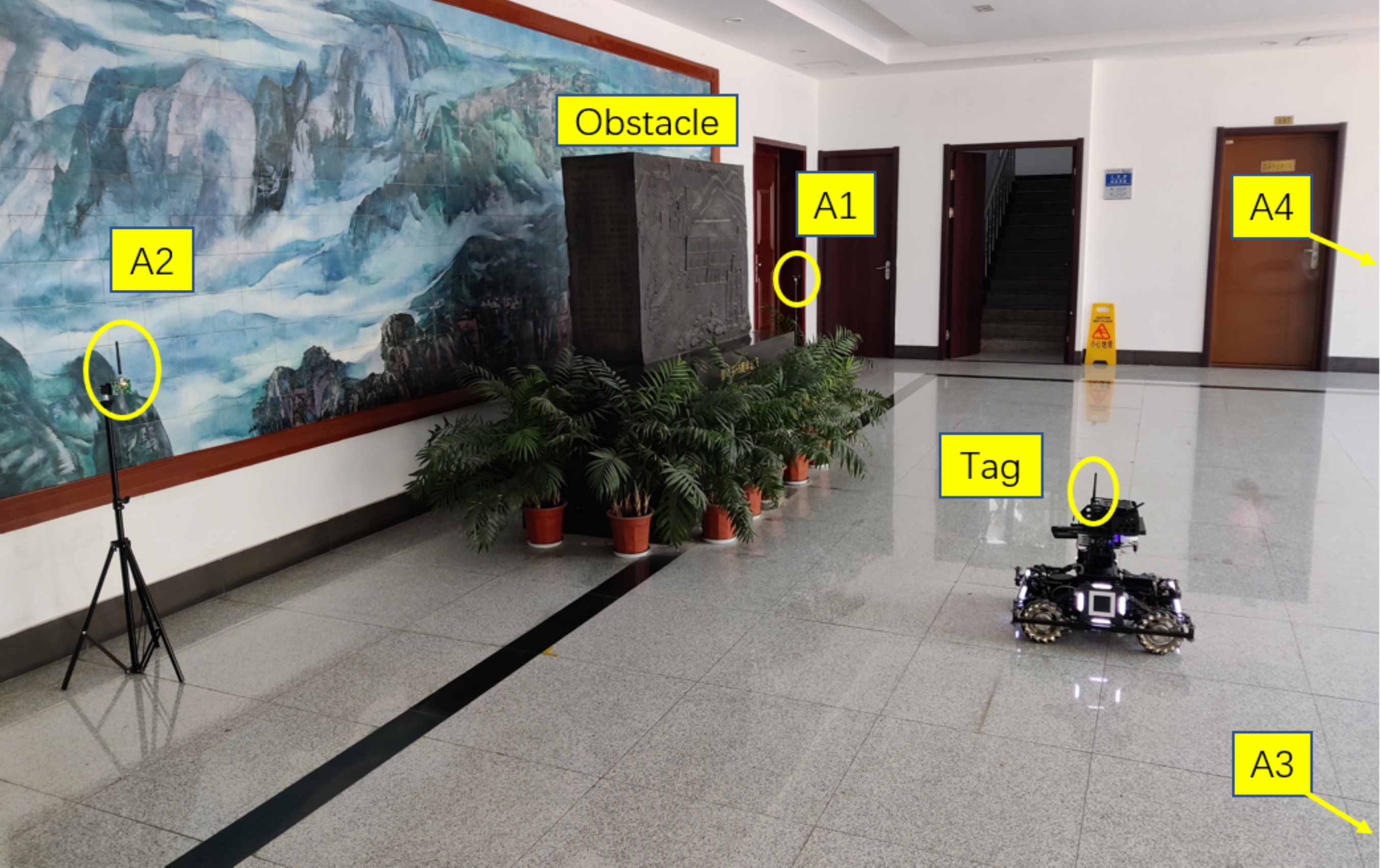}}
	\vspace{0mm}
	\subfloat[]{\includegraphics[width=8cm, height=5.2cm]{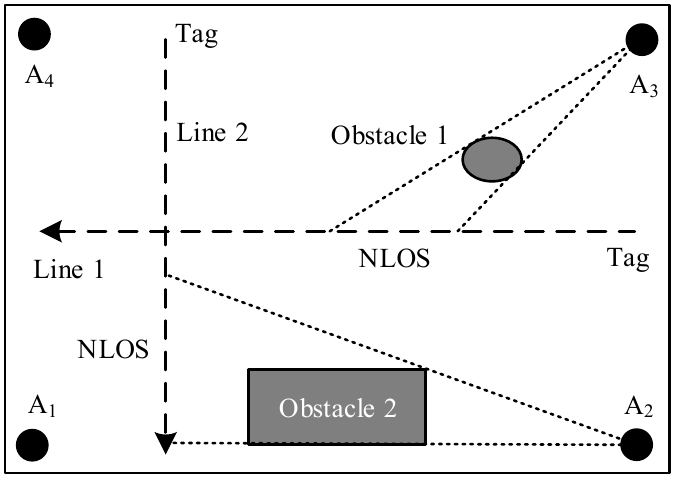}}
	\caption{The experiment setting. (a) Photo. (b) Schematic. For Line 1, a person obstructs the signal path between the anchor and the tag. For Line 2, the statue in the lobby is used as an obstacle.}
	\label{experiment setting}
\end{figure}

\begin{table*}[!b]
	\small
	\caption{Location Error(cm) for Simulations\label{simulation_RMSE}}
	\centering
	\begin{tabular}{p{2cm}<{\centering} p{1.2cm}<{\centering}p{1.2cm}<{\centering}p{1.2cm}<{\centering}p{1.2cm}<{\centering}p{1.2cm}<{\centering}p{1.2cm}<{\centering}p{1.2cm}<{\centering}p{1.2cm}<{\centering}}
		\toprule
		\multirow{2}{*}{Algorithm} & \multicolumn{2}{c}{Case 1} & \multicolumn{2}{c}{Case 2} & \multicolumn{2}{c}{Case 3} & \multicolumn{2}{c}{Case 4}\\
		~ & RMS & 90\% & RMS & 90\% & RMS & 90\% & RMS & 90\%\\
		\midrule
		LS      & 41.2 & 61.6 & 44.7 & 65.9 & 78.5 & 137.6 & 68.3 & 115.2\\
		RKF     & 40.4 & 59.7 & 43.8 & 64.4 & 79.6 & 140.2 & 69.1 & 116.4\\
		WLS-RKF & 1.7  & 2.1  & 1.9  & 2.0  & 1.9  & 3.3   & 1.8  & 3.0\\
		\bottomrule
	\end{tabular}
	\label{simulation table}
\end{table*}

We run 20 simulations for each case. Fig. \ref{simulation cases} shows the trajectories of Case 2 and 4. From Fig. \ref{simulation cases}(a), we can see that the trajectories are obviously biased due to NLOS when the tag passes by the obstacle, and the position error increases as more paths to the anchors are obstructed. RKF slows down the error growth but the trajectory still deviates gradually from the ground truth. The proposed WLS-RKF effectively mitigates the NLOS error and achieves accurate location results.
From Fig. \ref{simulation cases}(b), we can see that even the initial position A is biased due to NLOS. 
As the tag moves aside the obstacle, the paths to the anchors changes from LOS to NLOS or otherwise. WLS-RKF would adjust the NLOS judgement as the change occurs. The filtered trajectory "converges" to the true position at B, indicating that the NLOS identification is right and the error is migigated successfully.
From then on, the position results are always accurate.

We calculate the root-mean-square (RMS) and 90\% location error, as shown in
Table \ref{simulation table}. For Case 3 and 4, we ignore results of the first lap since the filter does not converge at the beginning.
With the WLS-RKF, the position error is reduced by over 95\% in all cases compared to unweighted LS.

\section{Experiments and Results}

\subsection{Experiment setup}
We carry out experiments in the lobby of the laboratory building. 
As shown in Fig. \ref{experiment setting}(a), there are 4 UWB anchors located at 
$A_1$(0.63, 0), $A_2$(7.57, 0), $A_3$(7.57, 5.40) and $A_4$(0.63, 5.40).  
An Unmanned Ground Vehicle (UGV) with an UWB tag on the gimbal is controlled to move along a line at the speed of about 0.5m/s. 
Each anchor or tag includes a DW1000 and an STM32F103 microcontroller. DW1000 is a low-power low-cost UWB transceiver. Through double-sided two-way ranging (DS-TWR) \cite{Decawave2017}, distances between the tag and the anchors can be measured from the TOF of UWB signals, and calibrated with the method in \cite{gonzalez2009mobile}. Preliminary experiments shows that the ranging interval is about 0.05s and the standard deviation measurement noise is about 2cm.

\begin{figure*}[tb]
	\centering
	\subfloat[]{\includegraphics[width=8.5cm]{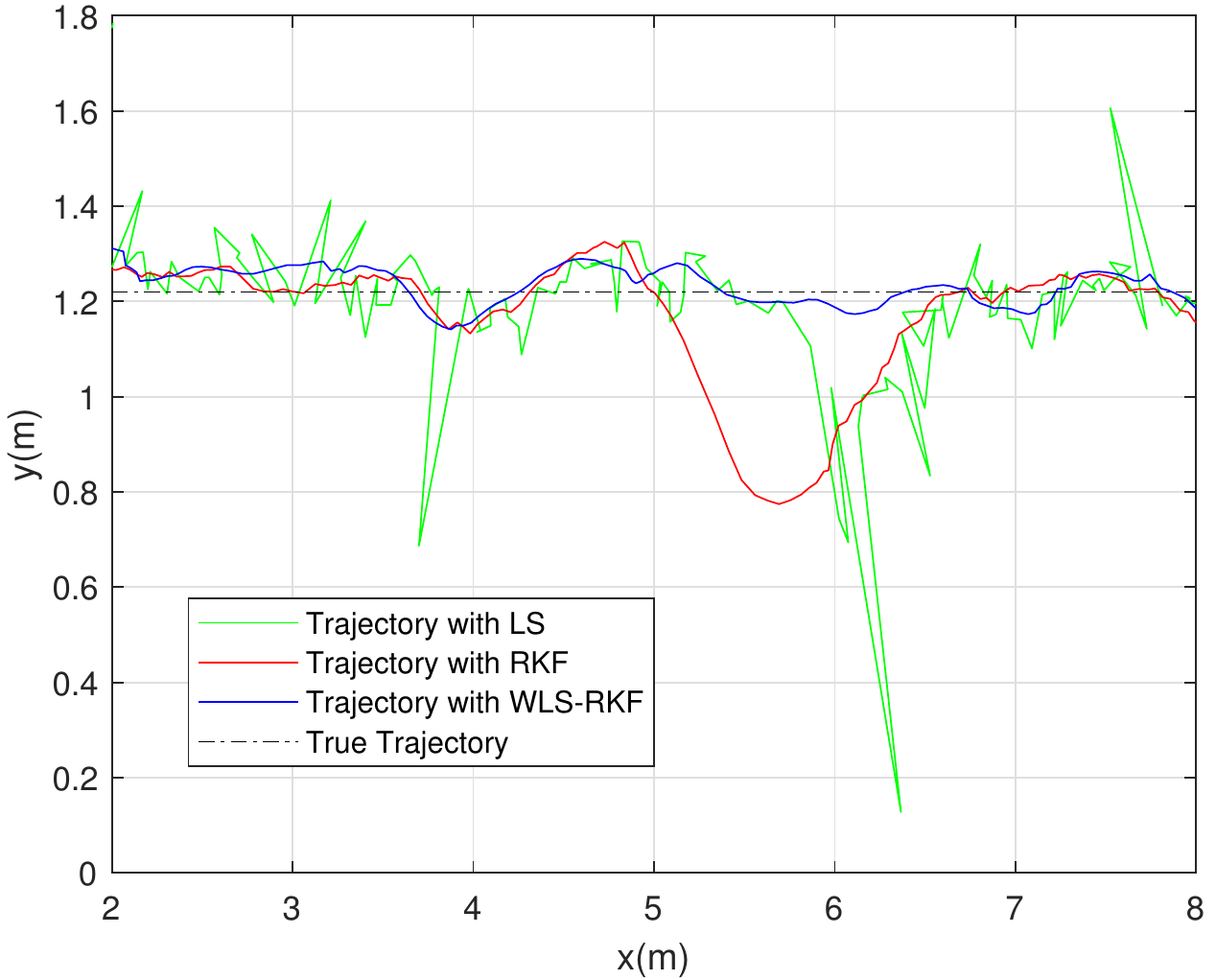}}
	\hfil
	\subfloat[]{\includegraphics[width=8.5cm]{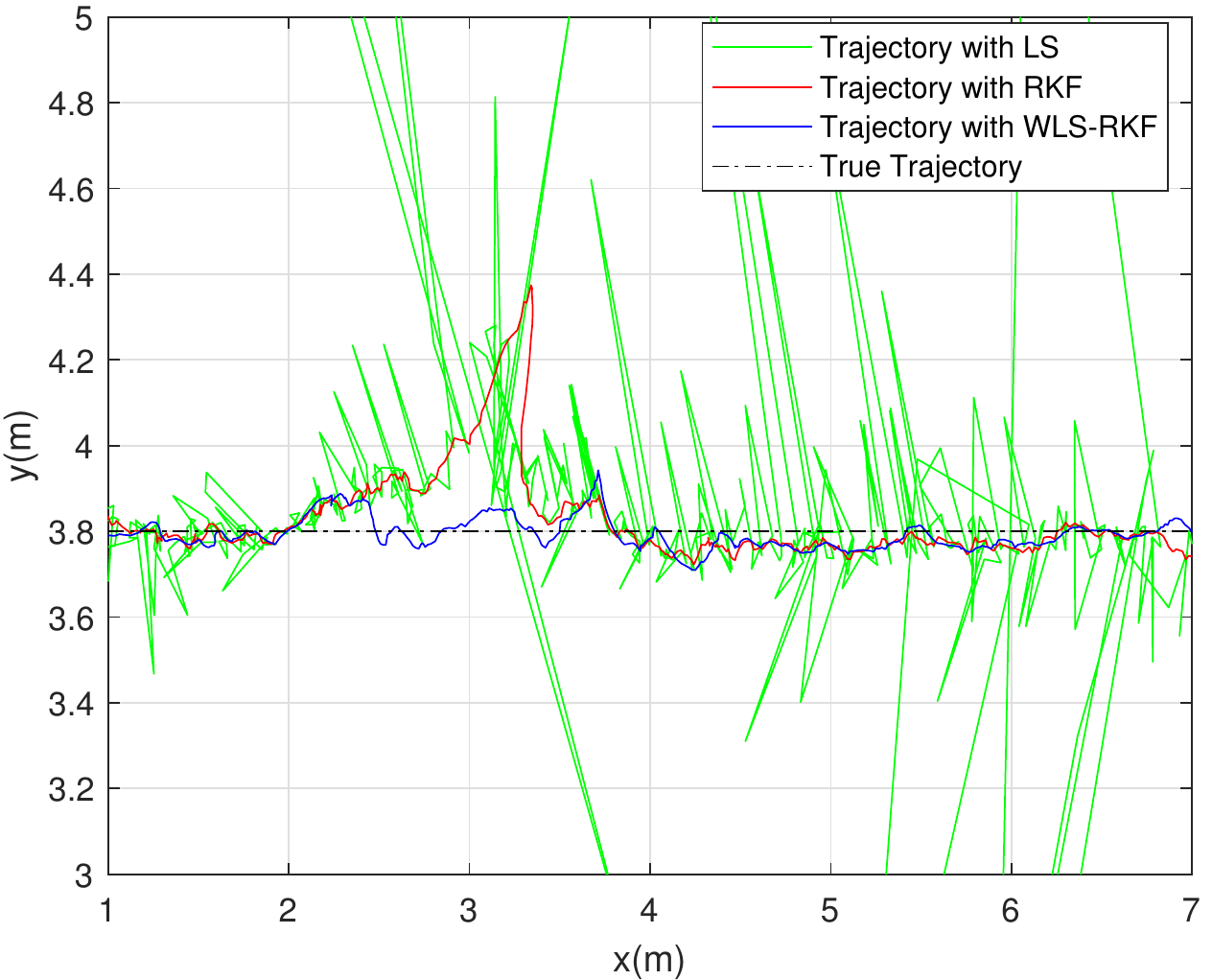}}
	\vspace{0mm}
	\subfloat[]{\includegraphics[width=8.5cm]{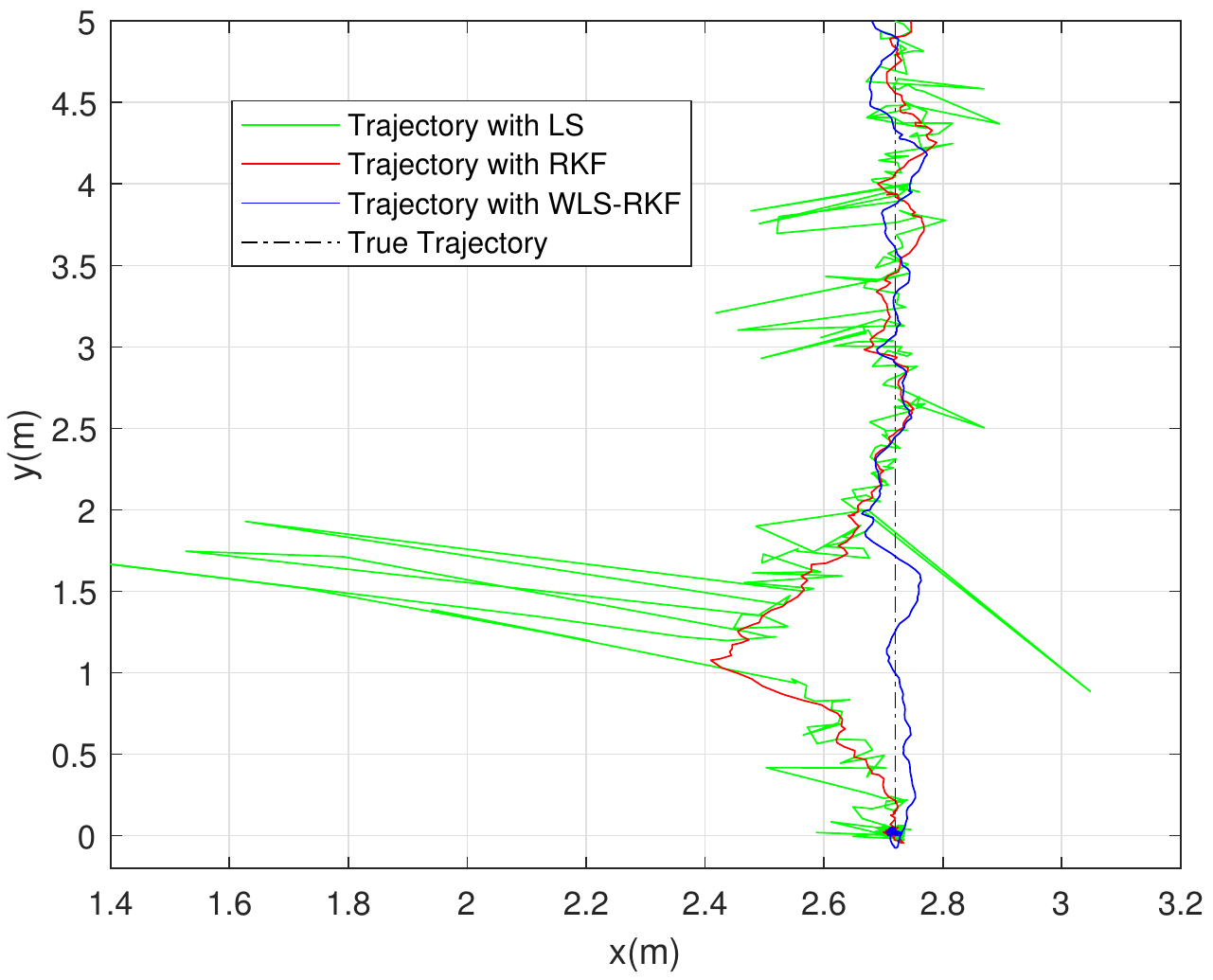}}
	\hfil
	\subfloat[]{\includegraphics[width=8.5cm]{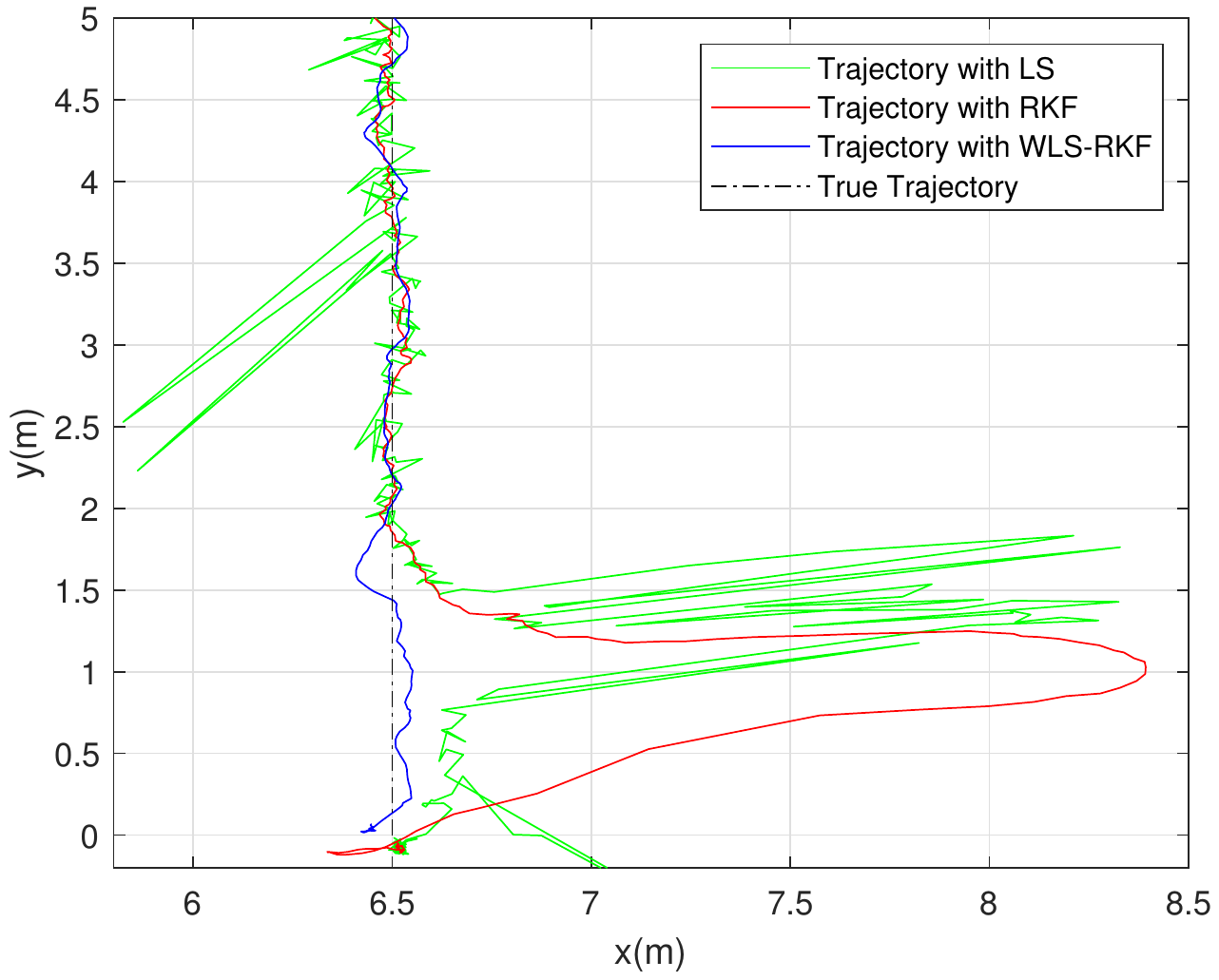}}
	\caption{Experiment results of LS, RKF and the proposed WLS-RKF. (a) Line 1-1. (b) Line 1-2.  (c) Line 2-1. (d) Line 2-2.}
	\label{experiment}
\end{figure*}

We consider 2 cases, where the UGV moves along Line 1 and Line 2, as shown in Fig. \ref{experiment setting}(b). Line 1 and 2 are parallel to the x-axis and y-axis respectively.
In the first case, a person stands inside the area as an obstacle. In the second case, the UGV passes by a statue, which introduces a great NLOS bias.

At every sample time, we collect the measured distances and apply the proposed WLS-RKF to get the position. All the parameters in the algorithm are the same as in the simulations. For comparison, the positions with unweighted LS and RKF are also calculated. 
The ground truth of the trajectory is measured by a Sokkia SRX1 total station, which has an accuracy of 1mm. 

\subsection{Experiment Results}

In each case, we perform experiments where the UGV passes on both sides of the obstacle, noted as Line 1-1, 1-2, 2-1 and 2-2.
Fig. \ref{experiment} shows the trajectories of each algorithm. The proposed WLS-RKF achieves a good performance in all the cases.
In comparison, the error grows more slowly with RKF but still accumulates as the UGV moves in NLOS condition. In Fig. \ref{experiment} (a) and (d), the filtered position is still biased after NLOS effect no longer exists. 

\begin{figure*}[tb]
	\centering
	\subfloat[]{\includegraphics[width=4cm]{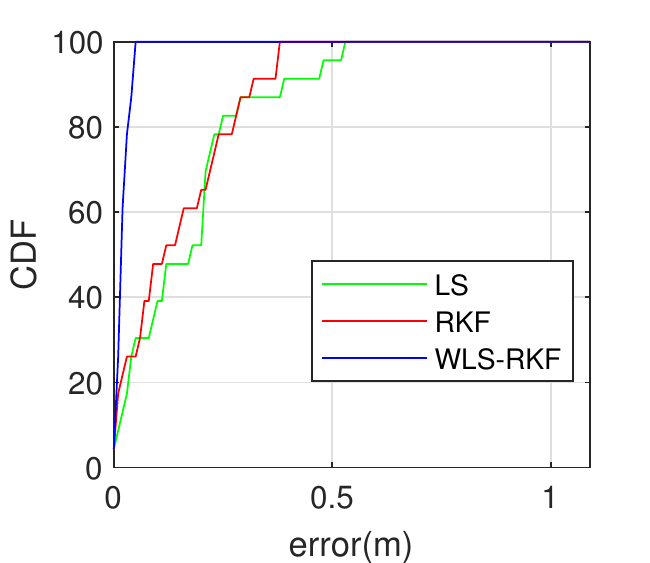}}
	\hfil
	\subfloat[]{\includegraphics[width=4cm]{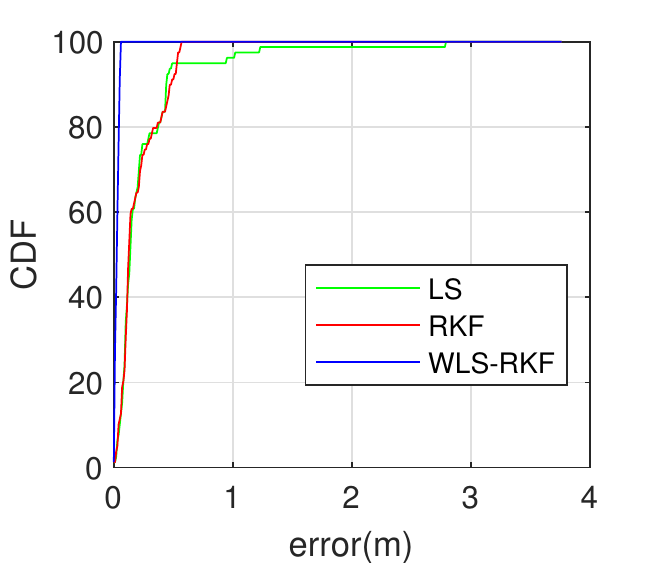}}
	\hfil
	\subfloat[]{\includegraphics[width=4cm]{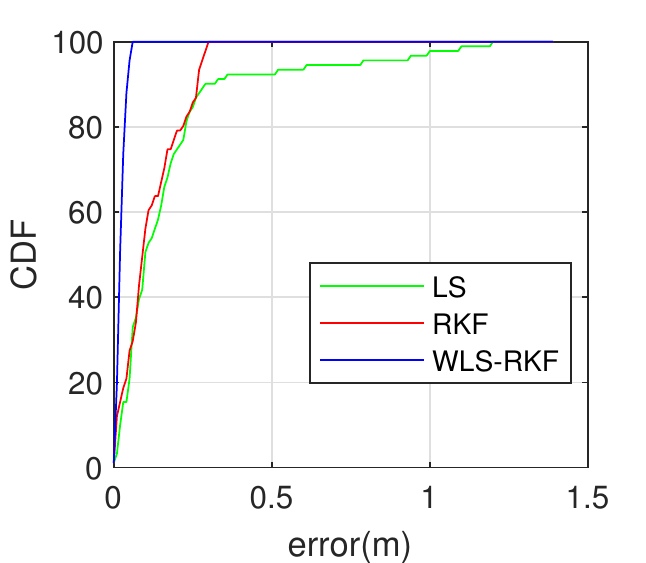}}
	\hfil
	\subfloat[]{\includegraphics[width=4cm]{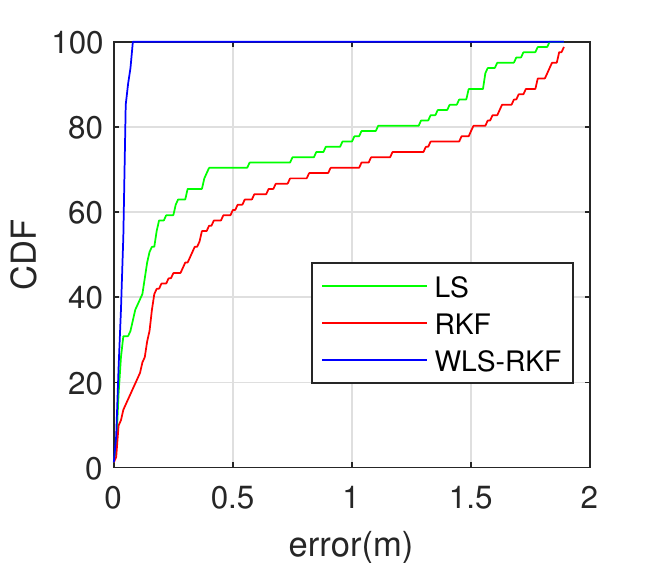}}
	\caption{CDF of position error for LS, RKF and the proposed WLS-RKF (a) Line 1-1. (b) Line 1-2.  (c) Line 2-1. (d) Line 2-2.}
	\label{experiment CDF}
\end{figure*}

Table \ref{experiment_error} shows the RMS and 90\% location error, and Fig. \ref{experiment CDF} shows the cumulative distribution function (CDF). As the trajectories are roughly parallel to the coordinate axes, we calculate the error in the y-axis for Line 1-1, 1-2, and in the x-axis for Line 2-1, 2-2.
Note that we only consider the section where the NLOS effect occurs, that is, $x \in (5.5, 7)$ for Line 1-1, $x \in (2, 4)$ for Line 1-2, and $x \in (0, 2)$ for Line 2-1 and 2-2. 
We can see that WLS-RKF mitigates the positioning error by over 90\% for all the cases. The RMS error is within 5cm.

\begin{table*}[tb]
	\small
	\caption{Location Error(cm) for Experiments\label{experiment_error}}
	\centering
	\begin{tabular}{p{2cm}<{\centering} p{1.2cm}<{\centering}p{1.2cm}<{\centering}p{1.2cm}<{\centering}p{1.2cm}<{\centering}p{1.2cm}<{\centering}p{1.2cm}<{\centering}p{1.2cm}<{\centering}p{1.2cm}<{\centering}}
		\toprule
		\multirow{2}{*}{Algorithm} & \multicolumn{2}{c}{Line 1-1} & \multicolumn{2}{c}{Line 1-2} & \multicolumn{2}{c}{Line 2-1} & \multicolumn{2}{c}{Line 2-2}\\
		~ & RMS & 90\% & RMS & 90\% & RMS & 90\% & RMS & 90\%\\
		\midrule
		LS      & 31.8 & 47.6 & 60.5 & 44.8 & 31.9 & 32.8 & 78.3 & 155.8\\
		RKF     & 20.6 & 37.5 & 26.1 & 50.1 & 14.7 & 26.2 & 94.6 & 177.8\\
		WLS-RKF & 2.5  & 4.4  & 3.1  & 5.1   & 2.7  & 4.1  & 4.1  & 6.3\\
		\bottomrule
	\end{tabular}
\end{table*}

\section{Conclusion} 
In this letter, we propose WLS-RKF, an NLOS mitigation method in UWB positioning. 
It leverages the measured distances and mitigates NLOS error using WLS technique and Kalman Filter without any prior knowledge about NLOS distribution or signal features. 
The simulation and experiment results show that WLS-RKF achieves good performances in various scenarios.

For future work, it may be possible for multiple sensor fusion. Besides UWB positioning, other sensors such as inertial measurement unit (IMU) and odometer can also be used to increase the robustness of the positioning system.

\bibliographystyle{IEEEtran}
\bibliography{IEEEabrv, ref}

\end{document}